\documentclass[9pt]{article}
\usepackage{spconf,amsmath,graphicx}
\usepackage{enumitem}
\setlist{nosep, leftmargin=14pt}
\usepackage{mwe} % to get dummy images

\usepackage{etoolbox}

\usepackage[table,xcdraw]{xcolor}
\usepackage{cite}
\usepackage{amsmath,amssymb,amsfonts}
\usepackage{algorithmic}
\usepackage{textcomp}

\usepackage{graphicx}       % graphics
\usepackage{multirow}
\usepackage{hyperref}
\title{ Attention-Aware Laparoscopic Image Desmoking Network with Lightness Embedding and Hybrid Guided Embedding}

\name{Ziteng Liu$^{1}$, Jiahua Zhu$^{2}$ , Bainan Liu$^{2}$, Hao Liu$^{3}$,  Wenpeng Gao$^{1,2,*}$, and Yili Fu$^{2,*}$ \thanks{$^{*}$Corresponding authors: Wenpeng Gao ({\tt\small wpgao@hit.edu.cn}) and Yili Fu ({\tt\small meylfu@hit.edu.cn})}}

\address{
$^{1}$School of Life Science and Technology,  Harbin Institute of Technology, Harbin, 150080, China \\
$^{2}$State Key Laboratory of Robotics and System,  Harbin Institute of Technology, Harbin,  150080,  China \\
$^{3}$State Key Laboratory of Robotics, Shenyang Institute of Automation, Shenyang, 110016, China}%

\begin{document}
\maketitle

\begin{abstract}

% This paper proposes a desmoking method for laparoscopic surgery smoke removal. As the surgical smoke is hetergeneous, a two-stage network is developed to estimate the smoke distribution and reconstruct the corresponding smoke-free scene with the hybrid embedding guidance, including the estimated smoke mask and the initial image. The network is designed following the concept of the atmospheric scattering model, which makes the network more explainable. To qualitatively and quantitatively evaluate the proposed method, a synthetic laparoscopic dataset is generated and a real laparoscopic dataset is obtained. Experimental results show that the Peak Signal to Noise Ratio (PSNR) of the proposed method is $1.9\%$ higher than that of the state-of-the-arts, while the run-time is shorter more than $40.9\%$. The proposed method is comparable to or even outperforms the state-of-the-arts in desmoking quality and computation cost.

{This paper presents a novel method of smoke removal from the laparoscopic images. Due to the heterogeneous nature of surgical smoke, a two-stage network is proposed to estimate the smoke distribution and reconstruct a clear, smoke-free surgical scene. The utilization of the lightness channel plays a pivotal role in providing vital information pertaining to smoke density. The reconstruction of smoke-free image is guided by a hybrid embedding, which combines the estimated smoke mask with the initial image.  Experimental results demonstrate that the proposed method boasts a Peak Signal to Noise Ratio that is $2.79\%$ higher than the state-of-the-art methods, while also exhibits a remarkable $38.2\%$ reduction in run-time. Overall, the proposed method offers comparable or even superior performance in terms of both smoke removal quality and computational efficiency when compared to existing state-of-the-art methods. This work will be publicly available on \href{http://homepage.hit.edu.cn/wpgao}{http://homepage.hit.edu.cn/wpgao}.}
\end{abstract}


\section{Introduction}
{Computer-aided minimally invasive surgery (MIS) is rapidly gaining popularity due to its ability to provide essential preoperative and intraoperative information. However, the presence of smoke during laparoscopic procedures can severely degrade the quality of laparoscopic images \cite{Alankar2016Joint}. \textcolor{black}{This degradation severely impacts subsequent sugical image processing steps, such as robotic visual navigation \cite{YANG2023104444} and soft tissue reconstruction \cite{YANG2023105989}.}  The smoke can obstruct the surgeon's view and impede visualization of the tissue, potentially compromising the accuracy of surgical assistance and even leading to serious complications. Thus, smoke removal task for laparoscopic images is highly desirable.}

{Existing methods for image desmoking (also referred to as dehazing)) are primarily based on the atmospheric scattering model (ASM) \cite{mccartney1977optics}, which utilizes handcrafted features and assumes that the predicted scene is a clear image \cite{ he2010single}.  It formulates a simple hypothesis on smoke images:
\begin{equation}\label{eq:asm}
    I\left ( x \right ) = J\left ( x \right )t\left ( x \right ) +A\left (1- t\left ( x \right )  \right ) 
\end{equation}
where $I(x)$ is the observed smoke image, $J(x)$ is the scene radiance to be recovered, $A$ is the global atmospheric light, $t(x)$ is the medium transmission. However, ASM-based desmoking methods may encounter limitations in complex scenarios, e.g., laparoscopic surgery \cite{Alankar2016Joint, chen2019smokegcn}.} 

{Recently, deep learning \cite{li2017aod,liu2019griddehazenet} has demonstrated significant potential in overcoming these limitations and achieving state-of-the-art performance in image desmoking.  The emergence of vision Transformer (ViT) \cite{dosovitskiy2020image} as a challenger to convolutional neural networks (CNN) has led to the development of ViT-based networks \cite{guo2022image,song2023vision}. Swin Transformer \cite{liu2021swin} is a modified ViT, which is designed to efficiently partition tokens into windows, enabling the computation of self-attention within each window while improving computing efficiency. However, the smoke generated during laparoscopic procedures is varied and does not depend on the depth of the tissue being operated on, effective smoke removal methods must possess the ability to detect smoke and prevent quality degradation in smoke-free regions \cite{chen2019smokegcn}. Therefore, supplementary information pertaining to the smoke can be provided for guidance.} 

\begin{figure*}[htb]
		\centering
		\includegraphics[width=\textwidth]{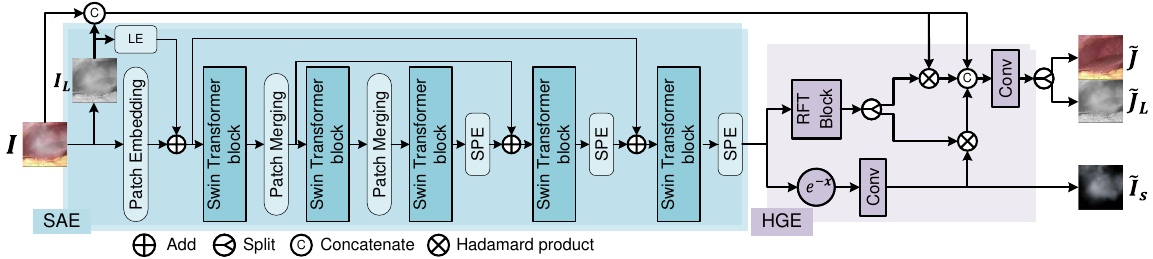}
		\caption{Visualization of the L-SAHGNet. The SAE predicts the smoke attention map from the input image $I$. The HGE takes the smoke attention map and $I$ as inputs, and predicts a smoke mask $\tilde{I}_s$ based on the ASM. Additionally, HGE produces a desmoked image $\tilde{J}$ with the guidance of both  $\tilde{I}_s$ and  $I$. $I_L$ and $\tilde{J}_L$ represent the initial lightness and the predicted lightness, respectively. Lightness is the L of HSL color model. The SPE is an up-sampling module. The RFT block transforms the smoke attention map into the weight matrices.}
		\label{fig:overview}
\end{figure*}

{In this paper, we proposed the L-SAHGNet,  an end-to-end deep learning network (Fig. \ref{fig:overview}) specifically designed for surgical smoke removal. L-SAHGNet takes smoke images as input and generates both the smoke-free image and the corresponding smoke mask as output.
The network consists of two modules: the Smoke Attention Estimator (SAE) and the Hybrid Guided Embedding (HGE). The SAE, resembling a U-Net-like Swin Transformer with a learnable lightness embedding, is responsible for estimating the attention map of the smoke present in the input image. In HGE,  the reconstruction process of the smoke-free image is guided by both the initial input image and the predicted smoke mask. To effectively preventing over-removal issues, the HGE utilizes the predicted smoke mask to impose strong constraints on the smoke removal area. }

\section{Method}

\subsection{Smoke Attention Estimator}

As illustrated in Fig. \ref{fig:overview}, the SAE comprises a 5-stage U-Net architecture, in which the Swin Transformer blocks serve as the basic units.

In the encoder, patch embedding and patch merging layers \cite{liu2021swin} are used for down-sampling the input image. Specifically, the patch embedding layer divides the input image into a sequence of patches and flattens each patch into a vector representation. These vectors are then processed by a linear layer to extract deeper features. Subsequently, the patch merging layer concatenates the patches and applies a linear layer to double the feature dimension while reducing the resolution by half. 

In the decoder, a soft patch expanding layer (SPE) is proposed for up-sampling. The SPE is a modified version of the patch expanding layer (PE) \cite{cao2022swin}. Unlike the PE layer that applies normalization on a single token, the SPE applies layer normalization on the entire feature map to keep the global attention strength information. The SPE is given as follows:

\begin{equation}
    SPE\left ( x \right ) = Flatten\left (LN\left ( PS\left ( {Conv} \left ( x \right )  \right )  \right ) \right )
\end{equation}
where LN is the layer normalization. PS is the pixel-shuffle operation. The flatten operation reshape the feature map to ViT tokens. The detailed parameter settings of SAE are listed in Table I. 

\begin{table}[!h]
\centering
\label{tab:setting}
\caption{Detailed SAE setting.}
\begin{tabular}{ll}
\hline
               & Setting               \\ \hline
Patch Size     & 4$\times$4            \\
Number of Blocks & {[}8,8,2,4,1{]}       \\
Embedding Dimension & {[}48,96,192,96,8{]} \\
Number of Heads  & {[}3,6,12,6,3{]}      \\ 
Size of SPE Conv Kernel & 1$\times$1, 1$\times$1, 3$\times$3\\ \hline
\end{tabular}
\end{table}

\subsection{Lightness Embedding Module}

\begin{figure}[!htbp]
    \centering
    \includegraphics{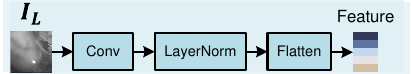}
    \caption{Structure of the lightness embedding module. }
    \label{fig:lightness_embedding}
\end{figure}

As the appearance of the smoke dramatically affects the dark channel \cite{he2010single}, the embedding of the channel-aware prior information can be involved in the position embedding \cite{dosovitskiy2020image} to improve the performance of ViT-based networks on desmoking tasks \cite{guo2022image}. 
In this work, the lightness, which is the L of the HSL color model, is taken as the channel-aware prior information and is embedded into the network through the proposed lightness embedding module.

As shown in Fig. \ref{fig:lightness_embedding}, the lightness embedding module (LE) proposed in this work can be treated as a learnable position embedding method. First, the lightness channel $I_L$ is calculated. Then, a convolution layer followed by a layer normalization is adopt to reduce the resolution and extract the deep features. The kernel size and stride are configured to match the patch size, ensuring that the resulting feature has the same shape as the patch token, which is the output of the patch embedding, after the final flatten operation.

\subsection{Hybrid Guided Embedding}

Inspired by the weak prior constraint used in \cite{song2023vision}, we introduce the predicted smoke information into the desmoking network and rewrite Eq. (\ref{eq:asm}) as:

\begin{equation}\label{eq:residual_prior}
    J\left ( x \right ) = I\left ( x \right ) + K\left ( x \right )I\left ( x \right ) + B\left ( x \right )I_{s}\left ( x \right )+ \varepsilon
\end{equation}
where $K\left ( x \right ) = 1/t\left ( x \right )-1$ and $B\left ( x \right ) = -1/t\left ( x \right )$, $I_{s}\left ( x \right )$ is the smoke mask,  $ \varepsilon$ is the intensity of the background image used to synthesize the smoke mask.

Thus, the structure of HGE straightforward yet comprehensible. 
By incorporating the texture prior $I\left ( x \right )$ within the network, it imposes a robust constraint on the image structure \cite{song2023vision}. The prior residual $K\left ( x \right )I\left ( x \right )$ serves as a signal-dependent noise that circumvents the vanishing gradient issue encountered in direct learning strategies. The guidance residual $B\left ( x \right )I_{s}\left ( x \right )$ introduces an extra \textit{a posteriori} smoke distribution into the residual component, facilitating the optimization of the smoke removal network through the provision of gradients.

\subsection{Restricted Field Transformation Block}

Due to the non-existence of the gradient of $1/t\left ( x \right )$ at t(x)=0, gradient explosion can occur when $t\left ( x \right )$ approaches zero. To address this issue and ensure numerical stability,  a field transformation ${t}\left ( x \right ) = e^{-\tilde{t}\left ( x \right )}$,  is employed.This transformation not only adheres to the definition of ASM but also effectively circumvents the aforementioned challenges, thus promoting a more robust and reliable computation.

Considering the exponential form of both $K\left ( x \right )$ and $B\left ( x \right )$ , directly performing the field transformation could potentially hinder the model's convergence. Hence, the widely utilized activation function, \textit{tanh}, which involves exponentiation operations, is employed to represent $1/t\left ( x \right )$. To counteract the reduction in range caused by this transformation, convolutional layers are positioned both before and after the activation function. The restricted field transformation block (RFT) is built as follow: 

\begin{equation}
\label{eq:rft_block}
    \left \{\tilde{K},\tilde{B}\right \}  = {Conv}_{1\times 1}\left (  tanh\left ( {Conv}_{1\times 1}\left ( \tilde{t} \left ( x \right )\right )   \right )  \right ) 
\end{equation}
where $\tilde{K}$ and $\tilde{B}$ are estimated $K\left ( x \right )$ and estimated $B\left ( x \right )$, respectively.

\subsection{Loss Function}

The loss of the L-SAHGNet is composed of the reconstruction loss and the guidance loss. The first term measures the mean squared error (MSE) between the predicted smoke-free image $\tilde{J}$ and the ground truth $J$, the second term is the MSE between the predicted smoke mask $\tilde{I}_s$ and its ground truth $I_s$, the third term is the MSE between the predicted lightness channel $\tilde{J}_L$ and its ground truth $J_L$. The loss function of the L-SAHGNet can be written as:

\begin{equation}\label{eq:lossfunc}
    L_{total} = L(\tilde{J},J) + \alpha_s L(\tilde{I}_{s},I_s) + \alpha_L L(\tilde{J}_{L},J_L)
\end{equation}
{where $L(A,B)$ is to calculate the mean squared error (MSE) of two inputs $A$ and $B$, $\alpha_s$ and $ \alpha_L$ are the weights to balance components and both empirically set to 1.0.}

\section{Experimental Results}

\subsection{Dataset}

The synthetic dataset, inspired by \cite{chen2019smokegcn}, was generated using Blender.  This dataset comprises synthetic smoke images that are composed of smoke-free images and smoke masks, which Blender generates employing 3 distinct density values.  For this study, a total of 18,140 smoke-free images were selected from the cholec80 dataset  \cite{twinanda2016endonet}, with 15,682 images from 21 videos designated for the training set and 2,458 images from 3 videos for the testing dataset. In all, the training set and the testing set contain 47,046 and 7,374 samples, respectively.  The real dataset includes 158 images sourced from 11 videos within the cholec80 dataset, which were not utilized in the synthetic dataset. All images in both datasets have been cropped to eliminate the black background and resized to 256 $\times$ 256 pixels for efficient training and validation processes.

\subsection{Implementation Details}

The proposed network was implemented using PyTorch and trained on a PC equipped with two NVIDIA 3090 GPUs. The training epoch was set to 300, and the initial learning rate was set at 1e-3. The variation of the learning rate was regulated using the cosine annealing strategy. AdamW was chosen as the optimizer for the training process.

\subsection{Quantitative  Evaluation}

\begin{table*}[ht]
		\centering
		\caption{Quantitative results on the synthetic and real dataset. Multiply-accumulate operations (MACs) were measured on 256$\times$256 images. $\uparrow$ and $\downarrow$ mean higher/lower score of this metric the better results methods achieve. \textbf{BOLD} and \textit{ITALIC} indicate the best and the second-best performance, respectively. }
		\resizebox{\textwidth}{!}{
\begin{tabular}{llllllllllll}
\hline
\multirow{2}{*}{Method} & \multicolumn{3}{c}{Synthetic}                       &  & \multicolumn{3}{c}{Real}                                  &  & \multicolumn{3}{c}{Overhead}     \\ \cline{2-4} \cline{6-8} \cline{10-12} 
                        & PSNR$\uparrow$ & SSIM$\uparrow$ & LPIPS$\downarrow$ &  & BRISQUE$\downarrow$ & NIQE$\downarrow$ & PIQE$\downarrow$ &  & \#Param & MACs   & Run-time (ms) \\ \hline
DCP                     & 21.51          & 0.880          & 0.093             &  & \textbf{20.32}               & 4.198            & \textbf{26.32}            &  & -       & -      & -         \\
AOD-Net                 & 22.13          & 0.887          & 0.076             &  & 22.60               & 4.580            & 37.70            &  & 0.002M  & 0.115G & 0.28          \\
GridDehazeNet           & 29.97          & 0.931          & 0.037             &  & 22.86               & 4.093            & 37.03            &  & 0.956M  & 21.49G & 7.11          \\
MSBDN                   & 26.34          & 0.958          & 0.027             &  & 22.66               & 4.202            & 36.36            &
  & 31.35M  & 41.54G & 11.39         \\
DehazeFormer-s          & \textit{32.16}          & \textbf{0.967}          & \textit{0.016}             &  & 21.36               & \textit{4.032}            & 27.90            &  & 1.283M  & 13.13G & 9.16          \\ \hline
Ours                    & \textbf{33.06}          & \textit{0.966}          & \textbf{0.013}             &  & \textit{21.02}               & \textbf{3.977}            & \textit{27.02}            &  & 4.814M  & 3.375G & 5.67          \\
Ours(w/o LE)            & 32.52          & 0.966          & 0.014             &  & 21.30               & 4.000            & 27.88            &  & 4.144M  & 3.310G & 5.58          \\
Ours(w/o SPE)           & 32.47          & 0.963          & 0.015             &  & 21.04               & 3.982            & 27.77            &  & 3.126M  & 3.174G & 5.31          \\
Ours(w/o RFT)           & 32.83          & 0.966          & 0.014             &  & 21.37               & 3.993            & 27.29            &  & 4.814M  & 3.375G & 5.69          \\
% Ours(w/o HGE)           & 31.60          & 0.949          & 0.025             &  & 20.79               & 4.165            & 23.58            &  & 4.764M  & 3.161G & 5.51          \\ 
\hline
\end{tabular}
		}

		\label{tab:results}
	\end{table*}

\begin{figure*}[!htbp]
		\centering
		\includegraphics[width=\textwidth]{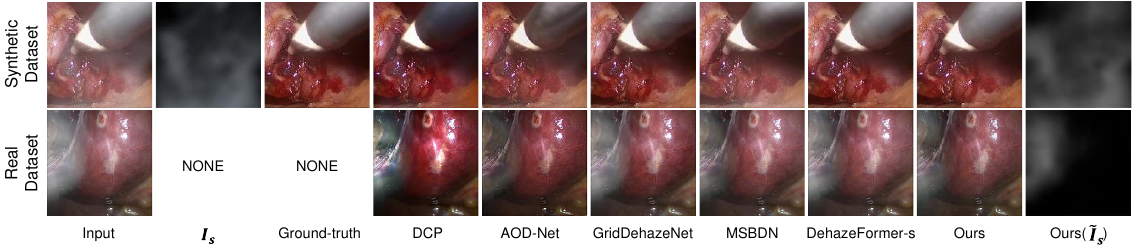}
		\caption{Comparison of the proposed method with the state-of-the-art methods on synthetic dataset and real dataset. $I_s$ is the ground truth of smoke mask.  }
		\label{fig:cmp_real_syn}
\end{figure*}

For a comprehensive comparison, we selected several state-of-the-art methods, namely DCP \cite{he2010single}, AOD-Net \cite{li2017aod}, GridDehazeNet \cite{liu2019griddehazenet}, and DehazeFormer \cite{song2023vision}, in this section. To ensure a fair comparison, all of these methods were trained using the synthetic dataset under the identical environments mentioned previously. Examples on synthetic dataset and real dataset are shown in Fig. \ref{fig:cmp_real_syn}, the proposed method is able to generate much clearer image and accurate smoke mask.

To evaluate the quality of the desmoked images, three quantitative evaluation metrics are employed: PSNR, SSIM and LPIPS \cite{zhang2018unreasonable}. As presented in Table \ref{tab:results}, the proposed method attains the highest scores for PSNR and LPIPS compared to all other methods. The PSNR of proposed network is $2.79\%$ higher than that of the second best method. However, in terms of SSIM, the score of L-SAHGNet is slightly lower than that of DehazeFormer-s. This decrease can be attributed to the inherent uncertainty associated with reconstruction using convolutional layer, which may result in a loss of high-frequency structural information.

The experimental evaluation on the real dataset involved three no-reference metrics: BRISQUE \cite{Mittal2012brisque}, NIQE \cite{Mittal2013niqe}, and PIQE \cite{Venkatanath2015piqe}. The proposed method, L-SAHGNet, demonstrated superior performance compared to other deep-learning-based methods across all three metrics, as shown in Table \ref{tab:results}. While the DCP method achieved the best results on BRISQUE and PIQE, it exhibited significant color distortion in its output images, as shown in Fig. \ref{fig:cmp_real_syn}. This finding underscores the ability of L-SAHGNet to generate more natural smoke-free images compared to other methods, emphasizing its effectiveness and superiority in handling real-world scenarios.

The run-time was recorded on the same machine, using the real dataset as inputs and with a batch size of 1 for all deep learning methods involved. The experimental results revealed that our method's run-time was only $61.8\%$ of DehazeFormer-s', which is the second best method in terms of PSNR. Furthermore, the proposed method achieved real-time performance, with a speed of 176.3 frames per second.

\subsection{Ablation Study}

\begin{itemize}

    \item {\textbf{Effectiveness of LE.} As shown in Table \ref{tab:results}, when the LE is deprecated, the network's performance considerably deteriorates on both the synthetic and real datasets. This result indicates that the lightness channel encapsulates the effective depth feature crucial for estimating the smoke distribution.}
    
    \item {\textbf{Effectiveness of SPE.} As shown in Table \ref{tab:results}, the network integrated with SPE layers consistently outperforms the one without SPE layers, which is constructed by replacing SPE layers with PE layers, on both the synthetic and real datasets. This enhanced performance could be attributed to the retention of global attention strength information.}

    \item {\textbf{Effectiveness of RFT.} As shown in Table \ref{tab:results}, the network integrated with RFT layers consistently outperforms the one without SPE layers, which is constructed by replacing $tanh$ with $e^x$, on both the synthetic and real datasets. These findings suggest that imposing range restrictions can contribute to more effective estimation of the weight matrix for residual terms, ultimately leading to improved network performance. }

    % \item Effectiveness of HGE. As shown in Table \ref{tab:results}, the performance of the network using HGE dramatically surpasses the network without HGE on most of the metric, e.g. the PSNR is increased $39.96\%$ and $6.8\%$ respectively, indicating the effectiveness of the HGE. 
 
\end{itemize}

\section{Conclusion}
% In this paper, we propose a real-time desmoking network for laparoscopic surgical smoke removing tasks. The network takes the initial images as prior knowledge to construct the residual terms and makes use of the smoke masks as posterior guidance. 
% The ASM is employed to establish constraints between the residual terms and make the network more explainable.
% Allowing for the lack of the data, a synthetic laparoscopic dataset is generated for training and validating the network and a real laparoscopic dataset is acquired for testing. The experimental results show the impressive performance of the proposed network on two datasets, which further proves the feasibility that the proposed network are capable of eliminating actual surgical smoke as long as the network trained on the synthetic laparoscopic dataset. It will remove the smoke and clean the scene for the robot in MIS.

This paper introduces a real-time desmoking network for laparoscopic surgery smoke removal. It uses initial image and corresponding lightness channel to construct residuals and smoke masks for guidance. The ASM establishes constraints between residuals, improving network explainability. The results show impressive performance on both datasets, proving the feasibility of the network to eliminate surgical smoke. 

\textcolor{black}{In the future, we will enhance the real dataset by adding more MIS scenarios and a comprehensive evaluation of the network performance will also be conducted.}

\section{Compliance with Ethical Standards}
This work is a study for which no ethical approval was required.

\section{Acknowledgments}

This work was supported by National Key R\&D Program of China (Grant No.2023YFB4705700), Self-Planned Task (NO. SKLRS202010B,  SKLRS202209B) of State Key Laboratory of Robotics and System (HIT).

Bibliography
\bibliographystyle{IEEEbib}  
\bibliography{desmoke}

\end{document}